\documentclass{emulateapj}
\journalinfo{To Appear in the Astrophysical Journal Letters}

\slugcomment{}
\shorttitle{X-ray Eclipses of U Scorpii 2010}
\shortauthors{D. Takei et al.}

\begin{document}
\title{X-ray Eclipse Diagnosis of the Evolving Mass Loss in the Recurrent Nova U Scorpii 2010}
\author{
D.~Takei\altaffilmark{1},
J.~J.~Drake\altaffilmark{1},
M.~Tsujimoto\altaffilmark{2},
J.-U.~Ness\altaffilmark{3},
J.~P.~Osborne\altaffilmark{4},
S.~Starrfield\altaffilmark{5}, \&
S.~Kitamoto\altaffilmark{6}
}
\email{dtakei@head.cfa.harvard.edu}
\altaffiltext{1}{Smithsonian Astrophysical Observatory,
                 60 Garden Street, Cambridge, MA 02138, USA}
\altaffiltext{2}{Japan Aerospace Exploration Agency, Institute of Space and Astronautical Science,
                 3-1-1 Yoshino-dai, Chuo-ku, Sagamihara, Kanagawa 252-5210, Japan}
\altaffiltext{3}{European Space Agency, XMM-Newton Observatory SOC, SRE-OAX,
                 Apartado 78, 28691 Villanueva de la Ca\~nada, Madrid, Spain}
\altaffiltext{4}{Department of Physics and Astronomy, University of Leicester,
                 Leicester, LE1 7RH, UK}
\altaffiltext{5}{School of Earth and Space Exploration, Arizona State University,
                 Tempe, AZ 85287-1404, USA}
\altaffiltext{6}{Department of Physics, Rikkyo University,
                 3-34-1 Nishi-Ikebukuro, Toshima, Tokyo 171-8501, Japan}

\begin{abstract}
 We report the \textit{Suzaku} detection of the earliest X-ray eclipse seen in the
 recurrent nova U~Scorpii 2010. A target-of-opportunity observation 15 days after
 the outburst found a 27$\pm$5\%\ dimming in the 0.2--1.0~keV energy band at the
 predicted center of an eclipse. In comparison with the X-ray eclipse depths seen
 at two later epochs by \textit{XMM-Newton}, the source region shrank by about
 10--20\%\ between days 15 and 35 after the outburst. The X-ray eclipses appear to
 be deeper than or similar to contemporaneous optical eclipses, suggesting the
 X-ray and optical source region extents are comparable on day 15. We raise the
 possibility of the energy dependency in the photon escape regions, and that this
 would be a result of the supersoft X-ray opacity being higher than the Thomson
 scattering optical opacity at the photosphere due to bound-free transitions
 in abundant metals that are not fully ionized. Assuming a spherically symmetric
 explosion model, we constrain the mass-loss rate as a function of time. For a ratio
 of actual to Thomson opacity of 10--100 in supersoft X-rays, we find a total ejecta
 mass of about 10$^{-7}$--10$^{-6}$~$M_{\odot}$.
\end{abstract}


\section{Introduction}
A nova explosion occurs in an accreting binary system comprising a white dwarf and a red
dwarf companion. When the amount of accreted material reaches a critical mass, hydrogen
fusion is triggered by a thermonuclear runaway on the white dwarf \citep{starrfield2008a}.
An event is mainly characterized by the development of photospheric emission powered by
nuclear burning after the explosion. This is first in the optical, where a sudden
increase in brightness is the result of radiative transfer through the optically-thick
ejecta. As the ejecta expand and become less opaque, the dominant emission shifts toward
higher energies, eventually becoming supersoft X-rays. The spectral hardening occurs
because the mass outflow diminishes with time and consequently the pseudo photosphere is
formed deeper in the expanding ejecta (e.g., \citealt{bath1978o}). For detailed reviews
see e.g., \citet{warner2003a,bode2008a}.

Our understanding of nova evolution to date has almost entirely resulted from
photometric and spectroscopic studies. While all novae are thought to be binaries,
and some fraction will inevitably be eclipsing, the utilization of eclipses during an
explosion as a powerful and direct probe of the emitting geometry has not yet been
fully realized. Observing such a phenomenon is difficult for most novae because they
spend many thousands of years between outbursts and are therefore observed only once,
with no prior information on the nature of the progenitor. However several novae have
been observed to explode more than once in a human lifetime and are called recurrent
novae. Since they can be studied in detail prior to an eruption, they are invaluable
for understanding the nature of novae and cataclysmic variables.

\medskip

In this letter, we investigate the development of the supersoft source (SSS) region
for the eclipsing recurrent nova U~Scorpii 2010. Using the \textit{Suzaku} satellite
we have detected a dimming 15 days after the outburst, corresponding to the earliest
post-outburst X-ray eclipse ever reported for this object. Through comparison with
a 30--50\%\ dimming in X-ray eclipses at two subsequent epochs observed by
\textit{XMM-Newton} \citep{ness2012a}, we find that the X-ray source shrank with
time. For the first time we are able to use direct geometric information provided by
supersoft X-ray eclipses to investigate the evolution of the radiatively-driven flow
in the early stage of an explosion.

\section{Target (U Scorpii)}\label{target}
U~Scorpii is an eclipsing binary undergoing nova explosions recurrently about every 10
years. Novae have been observed ten times, the most recent of which were in 1999 and
2010 \citep{schaefer2010f}. \citet{starrfield1988a} noted that the short time between
outbursts points to a mass-gaining white dwarf close to the Chandrasekhar limit. The
system has an inclination $\sim$80~deg, and the radii of the companion star and the
binary orbit are 2.7 and 6.9~$R_{\odot}$, respectively, according to
\citet{hachisu2000b}, and are 2.1 and 6.5~$R_{\odot}$ by \citet{thoroughgood2001m}.
The orbital period before the 1999 outburst was 1.2305521~d \citep{schaefer1995a},
while the period from 2001 to 2009 was 1.23054695~d \citep{schaefer2010e}, indicating
that it changed due to the 1999 eruption \citep{matsumoto2003a}. We adopt the
\citet{schaefer2010e} ephemeris with an origin of HJD~2451234.539. An optical eclipse
covers $\sim$18\% of an orbital phase in the quiescent state, which corresponds to
0.22~d \citep{schaefer2010e}.

\begin{figure}[tb]
 \epsscale{1.00}
 \plotone{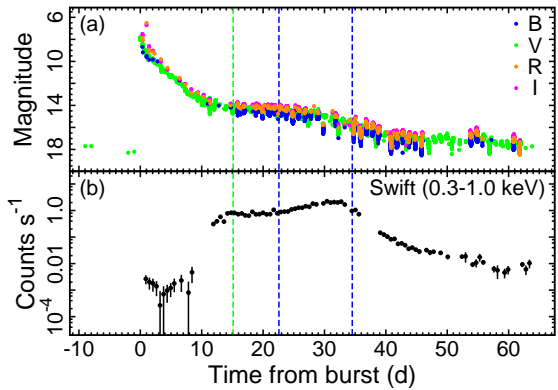}
 \caption{Development of (a) optical and (b) X-ray brightness in the nova outburst
 of U~Scorpii 2010. The origin of the abscissa is MJD~55224.4385 when the nova was
 discovered \citep{schaefer2010a}. The times of the \textit{Suzaku} and
 \textit{XMM-Newton} observations are shown by vertical dashed lines in green and
 blue, respectively. (a) Optical magnitudes are from the American Association of
 Variable Star Observers (AAVSO), Variable Star Observers League in Japan (VSOLJ),
 and table~2 in \cite{schaefer2011a}. (b) Background-subtracted count rates of the
 \textit{Swift} \citep{schaefer2010c} in the 0.3--1.0~keV energy band.
 }\label{figure:lcurve_optical}
\end{figure}

The tenth observed eruption was discovered on 2010 January 28.4385~UT (MJD 55224.4385)
by B.\,G.\,Harris \citep{schaefer2010a}. We define the epoch of the discovery as the
origin of time. A worldwide collaboration was organized, and subsequent studies were
conducted by ground- and space-based telescopes (e.g.,
\citealt{osborne2010,schaefer2010d}). The development of optical and X-ray
brightness is shown in Figure~\ref{figure:lcurve_optical}. Optical eclipses imply the
source region was 4.1, 3.4, and 2.2~$R_{\odot}$ on days 15--26, 26--41, and 41--67,
respectively \citep{schaefer2011a}. The eclipse shapes were consistent with
a spherical source until day 26, with apparently more disk-like morphology at later
times. A 30--50\%\ dimming in X-rays during expected times of eclipse were found by
\textit{XMM-Newton} on days 23 and 35 \citep{ness2012a}, suggesting the size of the
X-ray source was comparable to the optical and orbital sizes. The X-ray light curve
on day 23 exhibited oscillations, which \citet{ness2012a} interpreted in terms of
a reforming accretion disk.

\section{Observations and Reduction}\label{observations}
We performed target-of-opportunity observations of U~Scorpii on 2010 February 6, 9, and
12 (9, 12, and 15 days after the outburst, respectively) with the \textit{Suzaku} X-ray
satellite. \textit{Suzaku} \citep{mitsuda2007} has an X-ray Imaging Spectrometer (XIS:
\citealt{koyama2007}) and a Hard X-ray Detector (HXD: \citealt{takahashi2007,kokubun2007}).
We concentrate on the XIS data on day 15 that show a clear eclipse. The day 9 data are
poor in statistical quality, and the light curve on day 12 exhibited substantial
stochastic variability that prevented unambiguous measurement of the eclipse signature.

The XIS is equipped with four X-ray CCDs at the foci of four co-aligned X-ray telescope
modules \citep{serlemitsos2007a}. Three of them (XIS0, 2, and 3) are front-illuminated
CCDs sensitive in the 0.4--12~keV energy band, while the remaining one (XIS1) is a
back-illuminated CCD sensitive in the range 0.2--12~keV. XIS2 and a part of XIS0 are not
functional and their data were excluded. The XIS was operated in the normal clocking
mode with an 8~s frame time.

Data were processed with pipeline version 2.4.12.27. Events were removed during South
Atlantic anomaly passages, when night-earth elevation angles were below 5$^{\circ}$, and
day-earth elevation angles were below 20$^{\circ}$. The net exposure time is 27~ks. For
data reduction, we used the HEASoft package version 6.10 and the calibration database
version xis20090925/xrt20080709.

\begin{figure}[tb]
 \epsscale{1.00}
 \plotone{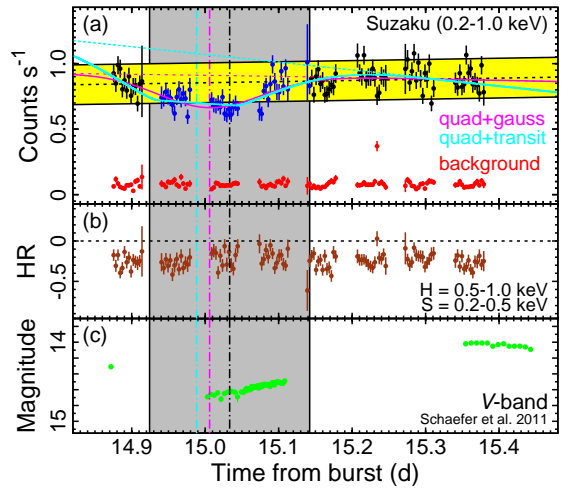}
 \caption{(a) XIS 0.2--1.0~keV count rates in eclipse (blue) and at other times (black),
 together with the background (red). The time of the optical eclipse is illustrated by
 the gray shaded region and the black dashed-and-dotted line. The best-fit quadratic
 model applied to the data outside eclipse is represented by a black dashed line. The
 yellow region shows the $\pm$90\% range of the black data points. The best-fit eclipse
 models are shown color-coded by solid lines; the quadratic component is illustrated by
 the dashed lines. The vertical dashed-and-dotted line in magenta and cyan indicate the
 Gaussian and transit center, respectively. (b) The hardness ratios (HRs) defined by
 (H$-$S)/(H$+$S), where H and S are rates in the 0.5--1.0~keV and 0.2--0.5~keV energy
 bands, respectively. (c) The \textit{V}-band magnitudes from \cite{schaefer2011a}.
 }\label{figure:lcurve_xis}
\end{figure}

\section{Analysis}\label{analysis}
X-ray light curves and spectra were constructed by taking source events accumulated from
a circular region of 130 pixels radius (2\farcm3) adaptively-chosen from XIS images to
maximize the signal-to-noise ratio. Barycentric correction of photon arrival times was
applied. Events taken with the three XIS were merged, and background was estimated from
an annular region with inner and outer radii of 180 and 250 pixels (3\farcm1--4\farcm2),
respectively. The photon pile-up fraction was $\lesssim$0.1\%. Simple spectral fitting
of the extracted photon events on day 15 yielded similar model parameters (blackbody
temperature $\sim$ 24~eV; Bremsstrahlung temperature $\sim$ 0.7~keV) to those obtained
by \citet{ness2012a} for the day 23 \textit{XMM-Newton} spectrum, indicating only slow
spectral evolution through this period. The SSS component dominates the X-ray spectrum,
in which the Bremsstrahlung component contributes at most only $\sim$1\%\ of the total
flux and/or count rates in the 0.2--1.0~keV energy band.

Effective areas were computed by incorporating the history of the satellite aspect
solution using a Monte Carlo simulation tool (\texttt{xissim}; \citealt{ishisaki2007})
that also accounts for an aspect-related vignetting of count rates by 10--20\%\ due to
the loss of a gyro system on \textit{Suzaku}. Finally, all count rates were normalized
to values corresponding to extraction from a circular aperture of 3$\arcmin$ radius.
The resulting background-subtracted light curve and hardness ratios in the 0.2--1.0~keV
energy band, together with the \textit{V}-band magnitudes from \citet{schaefer2011a},
are shown in Figure~\ref{figure:lcurve_xis}.

A shallow dip in the X-ray light curve is coincident with the predicted optical eclipse
on day 15 (Figure~\ref{figure:lcurve_xis}a). Comparison of the distribution of counts
during the optical eclipse with those outside of eclipse yields a null hypothesis
probability that the X-ray dimming is explained by a random source fluctuation on top
of the global trend of $<$0.01. No significant correlation was found between the flux
variation and the spectral hardness (Figure~\ref{figure:lcurve_xis}b), suggesting that
the variation of intrinsic colors does not affect the dimming. We thus conclude that
the X-ray dimming is due to the binary eclipse.

The X-ray eclipse was measured by fitting the entire light curve using a quadratic
function plus a Gaussian component to represent the source emission and approximate the
eclipse dimming, respectively (Figure~\ref{figure:lcurve_xis}a). The model yielded an
eclipse depth of 27$\pm$5\% at a center of 15.006$\pm$0.012 days after the outburst with
a duration of 0.17$\pm$0.04~d (0.13$\pm$0.03 orbital phase) in the full width at half
maximum (1$\sigma$ errors). The eclipse center was shifted by $\sim$0.03~d in comparison
with the optical ephemeris of \citet{schaefer2010e}. At face value this is a significant
offset. We note, however, that the optical eclipse minima observed by
\citet{schaefer2011a} show jitter of a similar magnitude. This jitter is likely due to
small fluctuating brightness inhomogeneities and departures from perfect sphericity in
the emitting region.

\section{Discussion}\label{discussion}
\subsection{Source Emitting Region}\label{analysis_eclipse}
The X-ray eclipse provides a means for constraining the geometry of the emitting region.
Assuming a spherically-symmetric source with no limb-darkening and the system parameters
of \citet{hachisu2000b}, the 27$\pm$5\%\ dimming on day 15 implies that the X-ray source
had roughly four times the area and twice the radius of the companion star, corresponding
to a source radius of 5.1$\pm$0.6~$R_{\odot}$. This is consistent with a fit to the
light curve using a quadratic function and analytic transit model \citep{mandel2002a}
assuming the same geometry, for which we obtained a radius 4.5$\pm$0.2~$R_{\odot}$ and
center epoch 14.988$\pm$0.007 days post-outburst; see Figure~\ref{figure:lcurve_xis}a.
By adopting the slightly more compact system parameters of \citet{thoroughgood2001m},
radius estimates are smaller by a factor of $\sim$0.8. This emitting region within the
residual outflow is large in comparison to the underlying white dwarf and was referred
to as a "corona" by \citet{ness2012a}.

The development of the source region can be studied by comparing the above radii with
the \textit{XMM-Newton} results taken at two different epochs \citep{ness2012a}. The
X-ray light curve on day 23 exhibited a dimming of up to $\sim$50\%\ with $\sim$20\%\
oscillations during an eclipse. The latter could be caused by absorption in a reforming
accretion disk \citep{ness2012a}, or by the geometrical eclipse of a photometrically
varying photon escape region caused by inhomogeneities and/or density instabilities
in the radiatively-driven flow (e.g., \citealt{shaviv2010o,shaviv2005e,owocki1988t}).
In contrast, the X-ray light curve on day 35 showed a clear eclipse with a $\sim$40\%\
dimming, corresponding to a source radius of $\sim$4.2~$R_{\odot}$, indicating that
the X-ray source shrunk by about 10--20\%\ between 15 and 35 days after the outburst.
This is the first quantitative X-ray measurement of the shrinking of the source radius
of an emerging SSS in a nova outburst.

The standard nova paradigm posits that the SSS emission emerges as ejecta expansion and
decline in the radiatively-driven flow allows photon escape at successively smaller
radial distance from a white dwarf surface until supersoft X-ray temperatures are
reached. Once this begins and the outer ejecta are thin, the source radius is of the
order of a few solar radii or smaller and the atmosphere is
possibly puffed-up and porous (e.g., \citealt{shaviv2010o}). This allows it to be much
larger than the Eddington luminosity and X-ray temperature would otherwise suggest
(e.g., \citealt{shaviv2005e}). A few thousand km/s outflow traverses the source in only
$\sim$100~s; at this point, any further change in radius probes the instantaneous
mass-loss rate in the outflow. We investigate this below, first assuming the simplest
ejecta model, and then examining the implications of an energy-dependent opacity by
comparing source radii in optical and X-rays on day 15.

\subsection{Simple Ejecta Model}\label{analysis_ejecta}
\begin{figure}[tb]
 \epsscale{1.00}
 \plotone{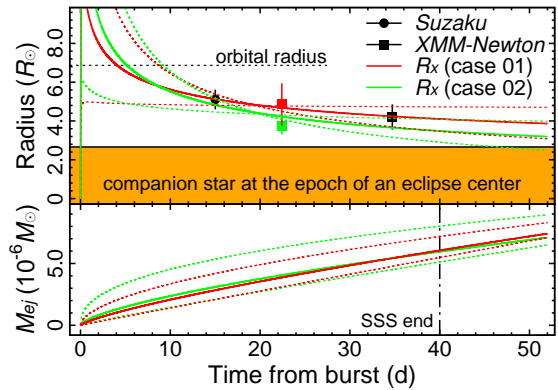}
 \caption{Top: The best-fit source radius model vs. time (solid) and 1$\sigma$
 confidence (dashed) for radii inferred from \textit{Suzaku} and \textit{XMM-Newton}
 eclipses. Models for day 23 eclipse depths of 30\%\ (red, case 1) and 50\%\ (green,
 case 2) are shown separately. The orange region and the dashed line indicate the
 sizes of the companion star and the orbital separation, respectively. Bottom:
 Total ejected mass vs. time corresponding to the upper panel models. The vertical
 dashed-and-dotted line shows the end of the SSS phase.
 }\label{figure:lcurve_radius}
\end{figure}

Assuming a spherically-symmetric flow from the white dwarf surface, the X-ray photon
escape radius $R_{x}$ is related to the optical depth by
\begin{equation}
 \tau{} = \int_{R_{x}}^{R_{out}}\kappa{}\rho{}dr, \label{eq:tau}
\end{equation}
where $R_{out}$ is the outermost radius of the ejecta, $\kappa$ is the ejecta opacity,
and $\rho{}$ is the density as a function of radial distance $r$. We define the X-ray
photon escape layer as a point at which $\tau{}$ is approximately unity. Assuming that
the ejected gases are accelerated within a small radial distance and that the terminal
velocity $v_{\infty{}}$ has not changed appreciably since the eruption (i.e., the time
scale of the terminal velocity decline is much larger than the time for the ejecta to
reach $R_{out}$), $R_{out}$ $\approx$ $v_{\infty}t$. We adopt $v_{\infty}$ $=$
3000~km~s$^{-1}$ from the estimate of \citet{yamanaka2010a} based on H$\alpha$ line
profiles, and $\kappa$ $=$ 0.4~cm$^{2}$~g$^{-1}$ as the Thomson scattering opacity in
a fully-ionized hydrogen-dominated gas. The radial density profile is given by
\begin{equation}
 \rho{} = \frac{1}{4\pi{}r^{2}}\frac{\dot{M}}{v_{\infty{}}}, \label{eq:rho}
\end{equation}
where $\dot{M}$ is the mass-loss rate. We next assume a time-dependent mass-loss rate
from the white dwarf given by (e.g., \citealt{bode2008a})
\begin{equation}
 \dot{M} = \dot{M}_{0}(t_{0}/t)^{p}, \label{eq:mdot}
\end{equation}
where $\dot{M}_{0}$ is an initial value normalized at time $t$ $=$ $t_{0}$ $=$ 1~s and
$p$ is a measure of the speed of its decline. Using these relations, we fitted the
observed radii of the X-ray source regions at three epochs (days 15 [\textit{Suzaku}],
23 and 35 [\textit{XMM-Newton}]) with two free parameters $\dot{M}_{0}$ and $p$. We
used the Gaussian estimate with the \citet{hachisu2000b} parameters on day 15 in order
to compare directly the eclipse depths with the other epochs. We further adopted 10\%\
uncertainties on the source radii derived by XMM-Newton based on the light curves in
\citet{ness2012a}. The source radius on day 23 depends on the interpretation of
the light curve oscillations, thus both cases of 30 and 50\%\ eclipse dimming (cases 1
and 2, respectively) were treated separately. The best-fit results and 1$\sigma$
confidence regions are shown in Figure \ref{figure:lcurve_radius}, where $\dot{M}_{0}$
$=$ 1$\times$10$^{-3}$ (6$\times$10$^{-5}$--3$\times$10$^{-2}$) $M_{\odot}$~yr$^{-1}$
and $p$ $=$ 0.23$\pm$0.22 in case 1; $\dot{M}_{0}$ $=$
5$\times$10$^{-3}$ (3$\times$10$^{-4}$--2$\times$10$^{-1}$) $M_{\odot}$~yr$^{-1}$ and
$p$ $=$ 0.33$\pm$0.25 in case 2.

Integration over the time-dependent mass-loss rate yields a rough estimate of the total
ejecta mass $M_{ej}$ as a function of elapsed time and is shown in
Figure~\ref{figure:lcurve_radius}. Assuming that the SSS faded and the mass outflow was
terminated on day 40 (Figure~\ref{figure:lcurve_optical}b), the total ejecta mass is
about 5--8$\times$10$^{-6}$~$M_{\odot}$ for the simple model. This is consistent with
the changing orbital period from long-term optical monitoring
(4.3$\pm$6.7$\times$10$^{-6}$ $M_{\odot}$: \citealt{schaefer2011b}), but slightly
higher than theoretical estimates for U~Scorpii (e.g., 4$\times$10$^{-7}$~$M_{\odot}$:
\citealt{starrfield1988a}; 2$\times$10$^{-6}$~$M_{\odot}$: \citealt{hachisu2000a}).
We here note that mass-related values also become smaller by the factor of $\sim$0.8 by
adopting \citet{thoroughgood2001m} parameters.

\subsection{Optical vs. X-ray Eclipses}\label{discuss_color}
\citet{schaefer2011a} found that the eclipse shapes in optical were consistent with a
spherical source with a radius of 4.1~$R_{\odot}$ on days 15--26. Since we have assumed
spherically-symmetric source regions whereas the optical eclipses \citep{schaefer2011a}
and the day 23 \textit{XMM-Newton} eclipses \citep{ness2012a} were accompanied by more
complicated structure, we cannot conclude with certainty that there is a significant
difference between the sizes of the optical and X-ray photon escape regions, though at
face value the different inferred radii suggest this on day 15 with the system
parameters of \citet{hachisu2000b}.

A remaining puzzle is the origin of the optical plateau started from $\sim$10--15 days
post-outburst. This optical flux clearly exceeds the extrapolated X-ray blackbody with
the Eddington luminosity of a Chandrasekhar white dwarf, though a blackbody is grossly
inadequate to describe the true spectrum and the magnitude of any optical excess
remains uncertain. \citet{hachisu2000b} argued the optical flux included reprocessing
photospheric emission on the surface of the reappearing accretion disk. Currently, no
clear evidence is found for this because it is unknown how early the accretion process
resumes after the explosion, though recent studies imply that the non-accretion case
is more likely, at least on day 15. Simulations of \citet{drake2010a} found that the
accretion disk was completely destroyed by the blast. \citet{mason2012u} and
\citet{worters2010a} suggest resumption of accretion based on spectroscopy and optical
flickering, respectively, around day 8, though the short term variability disappeared
on day 15 \citep{munari2010a}. \citet{schaefer2011a} found the optical source spherical
until day 26 based on eclipse mapping. \citet{ness2012a} further discussed reforming
accretion still on day 23. Alternatively, the reprocessing might occur in surrounding
ejecta with a substantial optical depth in supersoft X-rays, though it has to be noted
that any scenario at this time remains largely speculative.

If we assume that the optical flux on day 15 originated mostly from the surrounding
ejecta, the possible difference between the radii of optical ($R_{o}$ $\sim$
4~$R_{\odot}$) and X-ray ($R_{x}$ $\sim$ 5~$R_{\odot}$) sources can arise from the
energy dependence of the photon escape radii that would result from different optical
and X-ray opacities ($\kappa_{o}$ and $\kappa_{x}$, respectively). For a plasma at
temperatures of 10--100~eV,
the opacity in the supersoft X-ray range can be significantly larger than the Thomson
cross-section owing to abundant elements (primarily C, N, O, and Ne) not being fully
ionized. For a low-density plasma, the optical opacity will instead be close to the
Thomson value. Using the PINTofALE\footnote{http://hea-www.harvard.edu/PINTofALE/}
routine IONABS \citep{kashyap2000a}, we computed the cross-section assuming collisional
equilibrium at the temperature of 50~eV and found $\kappa_{x}$ $\sim$
50~cm$^{2}$~g$^{-1}$ in the 0.2--1.0~keV range, or $\sim$100 times the Thomson value.
The plasma in the ejecta is in fact likely to be photoionized rather than in collisional
equilibrium. For a constant wind velocity Eqns.~\ref{eq:tau} and \ref{eq:rho} imply for
optical depth unity $\kappa{}\rho{}r$ $\sim$ 1, or $\rho$ $\sim$ 5$\times$10$^{12}$
H~atoms~cm$^{-3}$ at 4~$R_{\odot}$. The ionization parameter, $\xi$ $=$ $L/n_{e}r^{2}$
(e.g., \citealt{tarter1969t}), is then of the order of 10$^{2}$ for the Eddington
luminosity and indicates the outflow is photon-dominated. Nevertheless, our
cross-section estimate suggets that the true supersoft X-ray opacity could be an order
of magnitude or more larger than that in the optical.

The density profile, and consequently the mass-loss rate and total mass loss, derived
from the X-ray eclipses scale inversely with the gas opacity, so that the total
mass-loss rate would be 5--8$\times$10$^{-6}$ $(\kappa_{o}/\kappa_{x})$~$M_{\odot}$.
We have argued that the X-ray opacity at the photosphere is higher than the Thomson
value that characterizes the optical range, and that this would give rise to the
observed difference in optical and X-ray source radii. Theoretical mass-loss estimates
are in the range 4$\times$10$^{-7}$--2$\times$10$^{-6}$~$M_{\odot}$
\citep{starrfield1988a,hachisu2000a}, indicating the true X-ray opacity might be
higher than the optical one by a factor of about 2--20. \citet{drake2010a} found the
initial explosion threw off probably no more than 10$^{-7}$~$M_{\odot}$ using
non-spherical hydrodynamic models and early X-ray luminosity constraints. Estimates
of the subsequent mass loss then imply that most of the mass was lost during later
evolution, confirming theoretical expectations (e.g., \citealt{gallagher1978t}).

\acknowledgments

We thank \textit{Suzaku} for undertaking our target-of-opportunity program, and
B.\,E.\,Schaefer for organizing the U~Scorpii 2010 collaboration.
Optical data are from AAVSO and VSOLJ databases. We acknowledge support from the
JSPS (D.\,T.), NASA contract NAS8-03060 to the CXC (J.\,J.\,D.), the UK Space
Agency (J.\,O.), and NSF and NASA grants to ASU (S.\,S.).


\end{document}